\documentclass[proc]{edpsmath}
\usepackage{amsfonts,amssymb,amstext}
\usepackage{amssymb,amsmath,babel,graphicx,epsfig} 
\usepackage[ruled]{algorithm}
\usepackage{algpseudocode}
\usepackage{booktabs}  
\usepackage[dvipsnames]{xcolor}
\usepackage{tikz}

\renewcommand{\Re}[0]{\mathbb{R}}
\newcommand{\PERTURBATION}[1]{#1'}
\newcommand{\SENSITIVITY}[1]{\bar{#1}}
\def\natexlab#1{{#1}}

\def\eq#1{{\begin{eqnarray*}#1\end{eqnarray*}}}

\def\p{\partial}
\def\d{\hbox{d}}

\newcommand{\beq}{\begin{eqnarray}}
\newcommand{\enq}{\end{eqnarray}}
\newcommand{\be}{\begin{eqnarray*}}
\newcommand{\en}{\end{eqnarray*}}

\newcommand{\R}{\mathbb R}

\newcommand{\E}{\mathbb E}

\newcommand{\ds}{\displaystyle}

\newcommand{\ii}{{\bf i}}

\newcommand{\bXn}{\bar X}
\newcommand{\bYn}{\bar Y}

\newtheorem{proposition}{Proposition} 

\def\d{\hbox{d}}

\def\redok#1{{#1}}

\newcommand{\esp}{\mathbb{E}}
\newcommand{\Esp}{\mathbb{E}}
\def\one{{\mathbf{1}}}
\def\CVA{\mbox{CVA}}
\def\LGD{\mbox{LGD}}
\def\Np{{N_{{\rm paths}}}}

\usepackage{listings}

\begin{document}

\title{Mini-Symposium on Automatic Differentiation and Its Applications in the Financial Industry}
%
\author{S\'ebastien Geeraert}\address{Murex (Paris, France).}
\author{Charles-Albert Lehalle}\address{Capital Fund Management (Paris, France) and Imperial College (London, U.K.).}
\author{Barak Pearlmutter}\address{Department of Computer Science, Maynooth University, Co.\ Kildare, Ireland.}
\author{Olivier Pironneau}\address{Sorbone Universite s, LJLL, UMR 7598, Case 187, 4 pl. de Jussieu, F-75252 Paris Cedex 5, France.}
\author{Adil Reghai}\address{Natixis (Paris, France).}

%
%
\begin{abstract}
Automatic differentiation \redok{has been} involved for long in applied mathematics as an alternative to finite difference to improve the accuracy of numerical computation of derivatives. Each time a numerical minimization is involved, automatic differentiation can be used. In between formal derivation and standard numerical schemes, this approach is based on software solutions applying mechanically the chain rule \redok{formula} to obtain an exact value for the desired derivative. It has a cost in memory and cpu consumption.

For participants of financial markets (banks, insurances, financial intermediaries, etc), computing derivatives is needed to obtain the sensitivity of \redok{their} exposure to well-defined potential market moves. It is a way to understand variations of their balance sheets in specific cases.
Since the 2008 crisis, regulation demands to compute this kind of
exposure to many different cases, to be sure market participants are
aware and ready to face a wide spectrum of market configurations. 

This paper shows how automatic differentiation provides a partial
answer to this recent explosion of computations to be performed. One part of the answer is a straightforward application of Adjoint Algorithmic Differentiation (AAD), but it is not enough. Since financial sensitivities involve specific functions and mix differentiation with Monte-Carlo simulations, dedicated tools and associated theoretical results are needed. We give here short introductions to typical cases arising when one uses AAD on financial markets.
\end{abstract}
\maketitle

\section{Introduction}

Numerous discussions with professionals underlined the recent importance of automatic differentiation in financial institutions.
It hence appeared natural to include a mini-symposium on this topic in the ``International Conference on Monte Carlo techniques''.

One of the roots of the recent interest of market participants for automatic differentiation is the ever demanding regulations on \emph{sensitivity} computations. For instance, Section B.2 and B.3 of Basel Committee on Banking Supervision's \emph{Minimum capital requirements for market risk} \cite{bis16riskb} contain a long list of sensitivities to be computed by   \redok{investment banks} on every product \redok{they own}. Sensitivities have then to be aggregated according to a regulatory formula to obtain the capital requirement the bank has to put in front of its risks.
Banks own financial products in their books, and the sensitivity of a financial product is the variation of its value with respect to a well defined variation of the market context. 
For instance, \cite[Section B.2.iii.]{bis16riskb} demands to compute \emph{delta}, \emph{vega} and \emph{curvature} for ``\emph{each financial product with optionality}'' owned by \redok{a} bank. These sensitivities are partial derivatives of the value of the instrument with respect to different \emph{risk factors}, grouped into \emph{risk classes}.
Seven risk classes are defined. As an example, the risk factors of the \emph{General Interest Rate Risk} class are \cite[pp20-22]{bis16riskb} :
\begin{itemize}
\item for the \emph{delta}: ``delta risk factors are defined along two dimensions: a risk-free yield curve for each currency in which interest rate-sensitive instruments are denominated and'' ten maturities.
\item For the \emph{vega}: ``within each currency, the vega risk factors are the implied volatilities of options that reference general interest rate risk-sensitive underlyings; further defined along two dimensions: \emph{Residual maturity of the underlying of the option at the expiry date of the option} and the \emph{Residual maturity of the underlying of the option at the expiry date of the option}.'' Each of them over 5 maturities.
\end{itemize}

We just listed more than 30 sensitivities to compute, belonging to one of the seven risk classes, to be applied on each financial instrument owned by the bank. Each sensitivity is a partial derivative.

Before this increase in regulatory demand, banks restricted computed sensitivities (i.e. partial derivatives) needed to hedge their portfolios, and almost no more. They used finite differences methods in the ``few'' needed directions.

The sudden increase of partial derivatives to be computed motivated the exploration other ways to compute them. Moreover, since regulatory updates continue to increase the list of the partial derivatives to compute, banks tried to identify methods allowing to compte new partial derivatives without changing their software code too much. 


With automatic differentiation (or adjoint algorithmic differentiation: AAD), you have not to develop a dedicated code each time you have a new product on which you need to compute sensitivities.
It is ``enough'' to embed your overall numerical libraries in a framework supporting AAD (see Section \ref{sec:sg}). 
Sessions of this mini-symposium (and hence sections of this paper) show it has an overall cost, but once this cost is paid, the marginal cost of computing more sensitivities is small.
ADD has hence been considered as one of the solutions to current regulatory demand and to be prepared to answer to future ones at a reasonable cost.
It explains why banks are carefully looking at ADD or are already using it to some extend.

If AAD exists for long and has numerous applications (see \cite{Baydin2015Adi} for a survey on interactions of AAD and statistical learning), the specificities of financial payoffs need some dedicated developments, as in \cite{PPS} (see Section \ref{sec:sg}). 
Moreover, the reformulation of some financial problems (like Credit
Value Adjustement, see Section \ref{sec:adil}) helps their inclusion in an AAD framework.

The sections of this paper reflect the sessions of the mini-symposium: each Section has been written by a speaker and contains a ``take home message'' about a specific aspect of AAD useful to use it on financial markets. The list of references at the end of the paper will help the reader to go deeper in this field.

In the first section, Barak Pearlmutter gives an overview of the most important aspects of AAD, in general. The reader can refer to \cite{Baydin2015Adi} for details. This section underlines the role of AAD in applied mathematics: progresses in numerical optimization, as a balance between manual derivation of specific functions involved in the criterion to minimize, and a pure numerical scheme via finite differences. Hence domains like non linear statistics, and more recently machine learning\footnote{Well-known machine learning frameworks, like Theano or TensorFlow, use AAD.}, took profit of progresses in AAD.

In Section \ref{sec:sg}, S\'ebastien Geeraert emphasizes some practical aspects he explored at Murex. 
He underlines how the specific mix of deterministic and stochastic processes involved in the payoffs of financial contracts can be exploited thanks to ad hoc choices to improve the efficiency of AAD. Typically ADD has to be used inside Monte-Carlo simulations, giving birth to specific constraints in terms of memory and cpu consumption.

In the next section, Olivier Pironneau put the emphasis on ways to apply AAD on non differentiable functions. This section follows the idea of applying the Vibrato method to a non differentiable function via smoothing. It allows to obtain accurate estimates for partial derivatives of order two in some cases.

Adil Reghai exposes in the last section how Natixis uses AAD in production to compute sensitivities of its exposure to potential market moves.
It demonstrates the advantage of AAD in the scope of the specific case of sensitivity to the counterparty risk. 


\section{What Automatic Differentiation Means to Me}%
\label{sec:barak}
\centerline{by \emph{Barak A.\ Pearlmutter}}
\bigskip

  We briefly (and from a self-serving biased and revisionist perspective) survey the basic ideas and intuitions of Automatic Differentiation: the efficient mechanical accurate calculation of derivatives of functions specified by computer programs.

\subsection{Introduction}

Automatic Differentiation or AD \cite{Griewank2008EDP} is a subdiscipline of numeric computation dedicated to the study of an enticing idea: the mechanical transformation of a computer program that computes a numeric function into an augmented computer program that also computes some desired derivatives accurately and efficiently.
The intuition for how this might be accomplished can be formed by considering a computer program implementing the function $f$ which maps an input vector to an output vector
\[
f : \Re^n \rightarrow \Re^m
\]
as consisting of a \emph{flow graph}: a directed acyclic graph leading from inputs to outputs whose edges carry $\Re$ values and whose nodes represent primitive arithmetic operations or fan-out. This flow graph is the natural ``split'' of the implementation of a computation in the usual atomic operators provided by the considered programming language. Automatic differentiation operates at the level of this graph, extending it to simultaneously computing the desired value, and the first term of its Taylor expansion.

\subsection{Forward AD}

\begin{figure}
  \fbox{
  \begin{tikzpicture}
    \node[shape=circle,draw=black] (mult) at (0,4) {$*$};
    \node[shape=circle,draw=black] (a) at (-5,5) {};
    \node[shape=circle,draw=black] (b) at (-5,3) {};
    \node[shape=circle,draw=black] (c) at (5,4) {};
    \path [->,line width=1.5pt](a) edge node[above] {\huge $a$} (mult);
    \path [->,line width=1.5pt](b) edge node[above] {\huge $b$} (mult);
    \path [->,line width=1.5pt](mult) edge node[above] {\huge $c$} (c);  

    \node[shape=circle,draw=black] (sincos) at (0,0) {sincos};
    \node[shape=circle,draw=black] (u) at (-5,0) {};
    \node[shape=circle,draw=black] (v) at (5,2) {};
    \node[shape=circle,draw=black] (w) at (5,-2) {};
    \path [->,line width=1.5pt](u) edge node[above] {\huge $u$} (sincos);
    \path [->,line width=1.5pt](sincos) edge node[above] {\huge $v$} (v);
    \path [->,line width=1.5pt](sincos) edge node[above] {\huge $w$} (w);
  \end{tikzpicture}
  }
  \fbox{
  \begin{tikzpicture}
    \node[shape=circle,draw=black] (mult) at (0,4) {$*$};
    \node[shape=circle,draw=black] (a) at (-5,5) {};
    \node[shape=circle,draw=black] (b) at (-5,3) {};
    \node[shape=circle,draw=black] (c) at (5,4) {};
    \node[shape=circle,draw=ForestGreen] (ta) at (-5,4.5) {};
    \node[shape=circle,draw=ForestGreen] (tb) at (-5,2.5) {};
    \node[shape=circle,draw=ForestGreen] (tc) at (5,3.5) {};
    \node[shape=rectangle,draw=blue] (jm) at (0,3.3) {
      $
      \left[\begin{array}{cc}
        b & a
      \end{array}\right]
      $};
    \path [->,line width=1.5pt](a) edge node[above] {\huge $a$} (mult);
    \path [->,line width=1.5pt](b) edge node[below] {\huge $b$} (mult);
    \path [->,line width=1.5pt](mult) edge node[above] {\huge $c$} (c);
    \path [->,line width=1.5pt,draw=ForestGreen](ta) edge node[below] {\color{ForestGreen}\huge $\PERTURBATION{a}$} (jm);
    \path [->,line width=1.5pt,draw=ForestGreen](tb) edge node[below] {\color{ForestGreen}\huge $\PERTURBATION{b}$} (jm);
    \path [->,line width=1.5pt,draw=ForestGreen](jm) edge node[below] {\color{ForestGreen}\huge $\PERTURBATION{c}$} (tc);

    \node[shape=circle,draw=black] (sincos) at (0,0) {sincos};
    \node[shape=circle,draw=black] (u) at (-5,0) {};
    \node[shape=circle,draw=black] (v) at (5,2) {};
    \node[shape=circle,draw=black] (w) at (5,-2) {};
    \node[shape=circle,draw=ForestGreen] (tu) at (-5,-0.5) {};
    \node[shape=circle,draw=ForestGreen] (tv) at (5,1.5) {};
    \node[shape=circle,draw=ForestGreen] (tw) at (5,-2.5) {};
    \node[shape=rectangle,draw=blue] (jsc) at (0,-1.4) {
      $
      \left[\begin{array}{c}
        w \\
        - v
      \end{array}\right]
      $};
    \path [->,line width=1.5pt](u) edge node[above] {\huge $u$} (sincos);
    \path [->,line width=1.5pt](sincos) edge node[above] {\huge $v$} (v);
    \path [->,line width=1.5pt](sincos) edge node[above] {\huge $w$} (w);
    \path [->,line width=1.5pt,draw=ForestGreen](tu) edge node[above] {\color{ForestGreen}\huge $\PERTURBATION{u}$} (jsc);
    \path [->,line width=1.5pt,draw=ForestGreen](jsc) edge node[above] {\color{ForestGreen}\huge $\PERTURBATION{v}$} (tv);
    \path [->,line width=1.5pt,draw=ForestGreen](jsc) edge node[above] {\color{ForestGreen}\huge $\PERTURBATION{w}$} (tw);
  \end{tikzpicture}
  }
\caption{Forward AD augments all primitive arithmetic operations in the original flow graph (top) to also propagate derivatives forward through the computation (bottom).}
\label{fig:forward}
\end{figure}
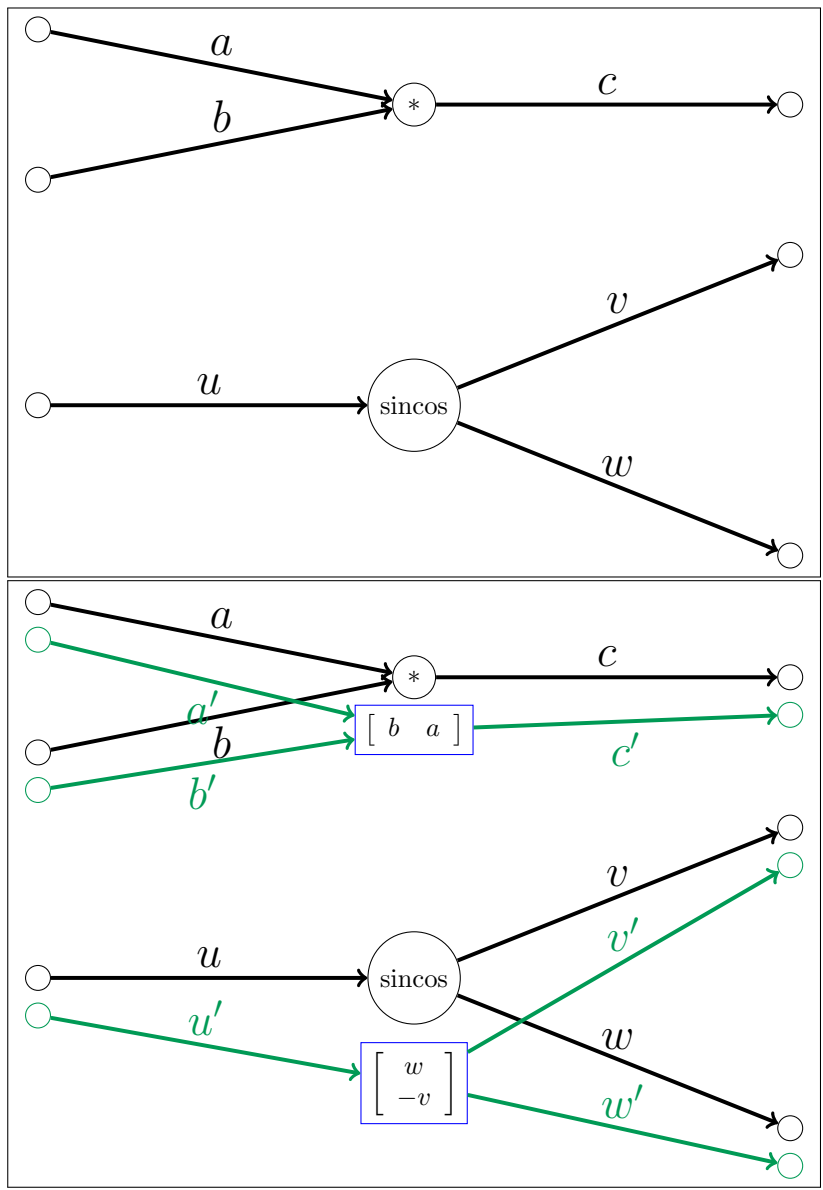

Using modern terminology, \cite{Wengert1964ASA} proposed that each real number $r$ in the flow graph could be replaced by a ``dual number'' \cite{Clifford1873}, a pair of real numbers $\langle r, \PERTURBATION{r} \rangle$ formally representing a truncated power series $r + \PERTURBATION{r} \epsilon + O(\epsilon^2)$. It will formally allow to carry not only operations on $r$ (and thus provide the standard operation), but on $r'$ too (providing the derivative in the direction of $r$).
Each operation in the dual space is made of a pair of operations: the standard one and its derivative. 
This means replacing each primitive arithmetic operation $g$:
\[
(v_1,\ldots, v_m) = g(u_1,\ldots,u_n)
\]
by its modification $Fg$, where the operator $F$ is a way to formalize we replaced $u$ by $\langle v, \PERTURBATION{v} \rangle$ in the expression of $g$:
\[
(\langle v_1, \PERTURBATION{v_1}\rangle,\ldots, \langle v_m, \PERTURBATION{v_m}\rangle) = Fg(\langle u_1, \PERTURBATION{u_1}\rangle,\ldots,\langle u_n, \PERTURBATION{u_n}\rangle)
\]
where
\[
\langle v_i, \PERTURBATION{v_i}\rangle = Fg_i(\langle u_1, \PERTURBATION{u_1}\rangle,\ldots,\langle u_n, \PERTURBATION{u_n}\rangle)
= \langle g_i(u_1,\ldots,u_n), \sum_j \PERTURBATION{u_j} (d/du_j) g_i(u_1,\ldots,u_n) \rangle.
\]
Figure \ref{fig:forward} shows the transformation of two basic operations (the upper box figures the flow graph for a multiplication and $\sin$ or $\cos$) into their dual representation.

Or, gathering values into vectors $\mathbf{u}=(u_1,\ldots,u_n)$, etc, and letting $J_g(\mathbf{u})$ be the Jacobian matrix of $g$ at $\mathbf{u}$, and gathering the primed values into vectors as well,
\begin{align*}
  \mathbf{v} &= g(\mathbf{u})
  &
  \PERTURBATION{\mathbf{v}} &= J_g(\mathbf{u}) \PERTURBATION{\mathbf{u}}
  &
  \langle \mathbf{v}, \PERTURBATION{\mathbf{v}} \rangle
  &=
  Fg(\langle \mathbf{u}, \PERTURBATION{\mathbf{u}} \rangle)
  =
  \langle g(\mathbf{u}), J_g(\mathbf{u}) \PERTURBATION{\mathbf{u}} \rangle
\end{align*}
It is now clear the dual space representation allows to simultaneously carry a operation and its derivative.

Assuming we have replaced each primitive arithmetic operation by an operation which does what the original operation did but also performs a Jacobian-vector product, as above, the entire computation will now perform a Jacobian-vector product.
We exhibit a small example of this transformation in Figure~\ref{fig:forward}, which graphically represents a procedure:
\begin{verbatim}
  function fig(double a, double b, double u) {
1:    double c = a * b;
2:    double[] (v,w) = sincos(u);
3:    return (c,v,w);
  }
\end{verbatim}
which is transformed into:
\begin{verbatim}
  function Ffig(double a, double da, double b, double db, double u, double du) {
1:    double c = a * b;
2:    double dc = a * db + da * b;
3:    double[] (v,w) = sincos(u);
4:    double dv = w * du;
5:    double dw = - v * du;
6:    return (c,dc,v,dv,w,dw);
  }
\end{verbatim}
or, if we bundle each primal variable \texttt{v} with \texttt{dv} into a structure \texttt{(v,dv)} stored in \texttt{Fv}, and for each primal subroutine or operator \texttt{fig} we use \texttt{Ffig} which calculates the primal but also propagated derivatives:
\begin{verbatim}
  function Ffig(double[] Fa, double[] Fb, double[] Fu) {
1:    double[] Fc = fTimes(Fa,Fb);
2:    double[][] (Fv,Fw) = Fsincos(Fu);
3:    return (Fc,Fv,Fw);
  }.
\end{verbatim}

Thanks to the implementation \texttt{Ffig} of the representation of routine \texttt{fig} in the dual space, it is now straightforward to calculate the full Jacobian by calling this $n=3$ times, each time calculating a single column of the Jacobian. It is enough to replace the second element of the dual variables by 0 or 1:
\begin{verbatim}
   ((c,j11),(v,j12),(w,j13)) = Ffig((a,1),(b,0),(u,0));
   ((c,j21),(v,j22),(w,j23)) = Ffig((a,0),(b,1),(u,0));
   ((c,j31),(v,j32),(w,j33)) = Ffig((a,0),(b,0),(u,1));
\end{verbatim}

This is ``Forward Accumulation Mode Automatic Differentiation'', or ``Forward AD''.  Note that since $n$ and $m$ are very small for primitive operations (one or two in general), the extra arithmetic added is a small constant factor overhead.  And note that we can allow the new ``prime'' values to flow through the computation in parallel with the original ones, meaning that the storage overhead is bounded by a factor of two.

\medskip

The same mathematical transformation as Forward AD has many names in
many fields: propagation of perturbations in machine learning, the pushforward in differential geometry, a directional derivative in multivariate calculus, forward error analysis in computer science, etc.
And many software tools exist that perform the Forward AD transformation, using a variety of implementation strategies ranging from source-to-source transformation (generally the fastest, but the least flexible or convenient) to overloading of operators in an object-oriented system (easy to implement and use but usually quite slow and sometimes not very robust due to semantic limitations of the overloading mechanisms.)

\subsection{Reverse AD}

\begin{figure}
  \fbox{
  \begin{tikzpicture}
    \node[shape=circle,draw=black] (mult) at (0,4) {$*$};
    \node[shape=circle,draw=black] (a) at (-5,5) {};
    \node[shape=circle,draw=black] (b) at (-5,3) {};
    \node[shape=circle,draw=black] (c) at (5,4) {};
    \node[shape=circle,draw=red] (da) at (-5,4) {};
    \node[shape=circle,draw=red] (db) at (-5,2) {};
    \node[shape=circle,draw=red] (dc) at (5,3) {};
    \node[shape=rectangle,draw=blue] (jm) at (0,3.3) {
      $
      \left[\begin{array}{cc}
        b & a
      \end{array}\right]
      $};
    \path [->,line width=1.5](a) edge node[above] {\huge $a$} (mult);
    \path [->,line width=1.5](b) edge node[above] {\huge $b$} (mult);
    \path [->,line width=1.5](mult) edge node[above] {\huge $c$} (c);
    \path [<-,line width=1.5,draw=red](da) edge node[left] {\color{red}\huge $\SENSITIVITY{a}$} (jm);
    \path [<-,line width=1.5,draw=red](db) edge node[left] {\color{red}\huge $\SENSITIVITY{b}$} (jm);
    \path [<-,line width=1.5,draw=red](jm) edge node[below] {\color{red}\huge $\SENSITIVITY{c}$} (dc);

    \node[shape=circle,draw=black] (sincos) at (0,0) {sincos};
    \node[shape=circle,draw=black] (u) at (-5,0) {};
    \node[shape=circle,draw=black] (v) at (5,2) {};
    \node[shape=circle,draw=black] (w) at (5,-2) {};
    \node[shape=circle,draw=red] (du) at (-5,-1) {};
    \node[shape=circle,draw=red] (dv) at (5,1) {};
    \node[shape=circle,draw=red] (dw) at (5,-3) {};
    \node[shape=rectangle,draw=blue] (jsc) at (0,-1.4) {
      $
      \left[\begin{array}{c}
        w \\
        - v
      \end{array}\right]
      $};
    \path [->,line width=1.5pt](u) edge node[above] {\huge $u$} (sincos);
    \path [->,line width=1.5](sincos) edge node[above] {\huge $v$} (v);
    \path [->,line width=1.5](sincos) edge node[above] {\huge $w$} (w);
    \path [<-,line width=1.5,draw=red](du) edge node[below] {\color{red}\huge $\SENSITIVITY{u}$} (jsc);
    \path [<-,line width=1.5,draw=red](jsc) edge node[above] {\color{red}\huge $\SENSITIVITY{v}$} (dv);
    \path [<-,line width=1.5,draw=red](jsc) edge node[below] {\color{red}\huge $\SENSITIVITY{w}$} (dw);
  \end{tikzpicture}
  }
\caption{Reverse AD augments all primitive arithmetic operations in the original flow graph to also propagate derivatives backwards through the computation.}
  \label{fig:reverse}
\end{figure}
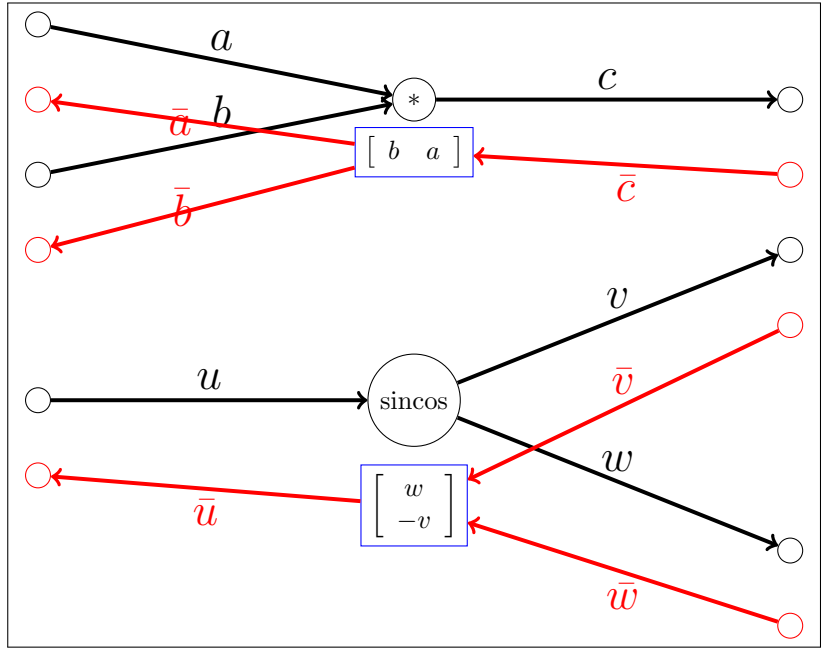

Consider an attempt to calculate the gradient of a function $f:\Re^n\rightarrow\Re$ using Forward AD.  This would require running the transformed function $Ff:\langle \mathbf{x},\PERTURBATION{\mathbf{x}} \rangle \mapsto\langle\mathbf{y},\PERTURBATION{\mathbf{y}}\rangle$ repeatedly, with $\PERTURBATION{\mathbf{x}}$ being cycled through the $n$ basis vectors $(1,0,0,\ldots,0), (0,1,0,\ldots,0), \ldots,  (0,0,\ldots,0,1)$.  When $n$ is large (say, $10^6$) this overhead is unacceptable.  Fortunately there is a different AD transformation which can calculate gradients more efficiently.  Each value $v$ in the original flow graph is augmented with an adjoint value $\SENSITIVITY{v}$, which are propagated backwards through the flow graph.  Each primitive arithmetic operation is modified (see Figure~\ref{fig:reverse}) to also multiply the vector of sensitivities of its \emph{output} by the \emph{transpose} of its Jacobian matrix, yielding the vector of sensitivities of its inputs.  Transforming the entire graph in this fashion allows the efficient computation of a Jacobian-transpose-vector product with only a small constant factor increase in operation count.  This allows the gradient to be calculated with only a small constant factor overhead!

We can view the computation of Figure~\ref{fig:reverse} as code instead of a graph.  Here, the primal must save some state for use in the reverse computation.  We do this explicitly, in a data structure.
\begin{verbatim}
  function Rfig(double a, double b, double u) {
1:    double c = a * b;
2:    double[] (v,w) = sincos(u);
3:    double[] RfigState = (a, b, u, c, v, w);
4:    return (c,v,w, RfigState);
  }
\end{verbatim}
The returned variable \texttt{RfigState} has to contain all that will be needed to compute the derivative of the function when only the back propagations of derivatives of its outputs of will be known. The function \texttt{Rfig} has to return, as usual, its three-dimensional result: $(c,v,w)$.
That is the purpose of the reverse phase: it just needs this ``state variable'' \texttt{RfigState} as first argument, and back propagated derivatives (i.e. derivatives of the outputs of the function, using the notation $\bar c, \bar v, \bar w$) as complementary arguments:
\begin{verbatim}
  function R2fig(double[] (a,b,u,c,v,w), double cbar, double vbar, double wbar) {
1:    double ubar = w * vbar - v * wbar;
2:    double bbar = a * cbar; 
3:    double abar = b * cbar;
4:    return (abar, bbar, ubar);
  }
\end{verbatim}
To understand these expressions, focus on $c = a\times b$: the first line of function \texttt{Rfig}. It is clear that if a function $\phi$ is applied to $c$, we have (with the notation ${\bar c} = \phi'(c)$, corresponding to a back propagation of the derivative coming from an application of functions to $c$):
$$  {\bar a}:= \frac{\partial \phi(a\times b)}{\partial a} =  b \times \phi'(c) = b\times{\bar c},$$
which is exactly the expression written an the third line of function \texttt{R2fig}.

This could be used to calculate the $m=3$ rows of the Jacobian by calling \texttt{R2fig} three times, each time calculating one row.
\begin{verbatim}
  (RfigState, c, v, w) = Rfig(a,b,u);
  (j11, j21, j31) = R2fig(RfigState, 1, 0, 0);
  (j12, j22, j32) = R2fig(RfigState, 0, 1, 0);
  (j13, j23, j33) = R2fig(RfigState, 0, 0, 1);
\end{verbatim}

\newcommand{\smallarray}[2]{%
\bigl[\begin{array}{c}\text{\footnotesize\(#1\)}\\[-1ex]\text{\footnotesize\(#2\)}\end{array}\bigr]%
}

Although low overhead for computing a gradient, in terms of operation count\footnote{For our example the gain is not huge, since the forward computation needs $3+1+2=6$ (see lines 2, 4 and 5 of \texttt{Ffig}) versus $3+2$ for the backward version (i.e. all computations in function \texttt{R2fig}), at the price of the storage in \texttt{RfigState} of all six variables involved in the primal expressions). But for more complex expressions the gain is more clearly at the advantage of the backward implementation, at a cost of more storage.}, makes this technique extremely useful, there are a variety of technical difficulties associated with Reverse AD.
Since the derivatives are propagated backwards relative to the original computation, the original values might need to be preserved, increasing the storage burden.
Fanout in the original computation corresponds to a linear primitive operation $\smallarray{v_1}{v_2}=\smallarray{1}{1} [u]$, implying that the adjoint computation must multiply by the transpose of this tiny matrix,
$\SENSITIVITY{u}=[1~1]\smallarray{\SENSITIVITY{v}_1}{\SENSITIVITY{v}_2} =\SENSITIVITY{v}_1+\SENSITIVITY{v}_2$,
corresponding to a simple addition. In other words, fanout in the original computation graph results in addition in the reverse-direction adjoint computation graph. 
These complexities of the transformation makes efficient implementation, via a source-to-source transformation of the code, quite involved.  The most popular implementation strategy is to actually store the dataflow graph and later retrace it in reverse order, which imposes considerable constant-factor overhead.

Although this transformation was formulated in the AD community, and automated, by \cite{Speelpenning80} and others \cite{Griewank-2012a}, it was known, albeit not automated, in the field of control theory where it was known as the Pontryagin Maximum Principle \cite{PONTRYAGIN61A}, in Mathematical Physics \cite{FEYNMAN39A} where it was known as an adjoint calculation, in Machine Learning where it is known as backpropagation \cite{Rumelhart1986LIR}, and in differential geometry where it is known as the pull-back.
The history of the discovery of both Forward AD and Reverse AD is fascinating, with many brilliant and colorful players. This former is discussed by \cite{Iri1991HoA}, and the latter, at some length, by \cite{Griewank-2012a}.

\subsection{Practical AD Systems and Applications}

Most useful complicated numeric routines are not so simple as a static dataflow graph, and part of the complexity of AD is handling control.  Other ``modes'' of AD are also available, to compute higher-order derivatives \cite{Karczmarczuk2001FDo, Pearlmutter2007LMH}, to vectorize multiple similar derivative calculations with stacked tangents or cotangents sometimes called vector mode, to conserve storage while still calculating gradients efficiently \cite{Griewank1992ALG, AD2016a}, and many many others.  Another substantial body of work has arisen as a consequence of the fact that taking exact derivatives and numeric approximation for implementation do not commute---a particularly difficult issue with loops that iterate to convergence.  And yet more work has gone into aggressive implementation technologies, with enormous efforts put into aggressive systems like TAPENADE \cite{Hascoet2004TUG, Pascual2005EoT, Pascual2008TfC}, ADIFOR \cite{Bischof1996UEw}, and others.
Currently, AD is enjoying substantial applications not just in finance \cite{Bischof2002ADf} and scientific computation like climate science \cite{Hasselmann1997SSo}, but also in machine learning \cite{Baydin2014ADo, Gelman2015SAP, Duvenaud-Adams-2015a}.
The field of AD is active, and welcoming to newcomers.
A catalog of AD implementations and literature is maintained at a community web portal, {\tt http://www.autodiff.org/}.

\subsection{An AD Dream}

In considering AD, it may be helpful to distinguish a hierarchy of notions, spanning a spectrum of automation and generality.
\begin{description}
\item[AD Type I] A calculus for efficiently calculating derivatives of functions specified by a set of equations.
\item[AD Type II] Rules applied manually which mechanically transform a computer program implementing a numeric function to also efficiently calculate some derivatives.
\item[AD Type III] A computer program which automatically transforms an input computer program specifying a numeric function into one that also efficiently calculates derivatives.
\item[AD Type IV] AD-taking operators included in language in which numeric algorithms are written, and they can be applied to procedure which use them.
\end{description}
The present author's AD research program, done in collaboration with Jeffrey Mark Siskind, has been to bring Type~IV into existence, and to make it fast,  robust, general and convenient \cite{Pearlmutter2008UPL}.
These efforts have therefore focused not just on formalizations of the AD process suitable for mathematical analysis and integration into compilers, but also in making AD more general and robust, allowing the AD operators to be used as first-class operators in a nested fashion, thus expanding the scope of programs involving derivatives that can be written succinctly and conveniently \cite{Radul-etal-2012a, Radul-etal-2012b, Baydin2015Adi, Baydin2015DAD}.
We have also explored the limits of speed that Type~IV AD integrated into an aggressive compiler can attain \cite{AD2016b}.

\bigskip

It is our hope that AD will become as ubiquitous for all numeric programmers as libraries of linear algebraic routines are now, and that optimizing compilers will in the future support AD with the same nonchalance with which they currently support trigonometric functions or loops.
\medskip

\paragraph{\bf Acknowledgments.}
This work was funded by Science Foundation Ireland grant 09/IN.1/I2637.
Thanks to collaborators Jeffrey Mark Siskind and Atilim Gunes Baydin for many helpful discussions.


\section{Practical implementation of adjoint algorithmic differentiation (AAD)}
\label{sec:sg}
\centerline{by \emph{S\'ebastien Geeraert}}
\subsection{Aim of the study}
\par In this study, we show a way of implementing adjoint algorithmic differentiation (AAD). AAD is a method producing the exact derivatives of a computation, in an automatic and efficient way. By using pathwise differentiation with AAD, we can compute Monte Carlo sensitivities of a price: if the value of the payoff of a financial contract is given by $p(\theta)=\esp\left[f(\theta,W)\right]$ where $\theta$ is a parameter and $W$ is a random variable we can estimate its derivative by 
$$p'(\theta)=\esp\left[\frac{\partial f}{\partial \theta}(\theta,W)\right]\approx \frac{1}{M} \sum_{i=1}^M \frac{\partial f}{\partial \theta}(\theta,W_i)$$
by swapping the differentiation and the expectation. The advantage of AAD is that the computation time does not depend on the number of sensitivities: by differentiating the operations in reverse (starting with the final computations and going back to the initial computations), we can progressively compute the influence of each intermediate result on the final output. Thus, we can compute every sensitivity at once. In the existing automatic differentiation libraries, there are two methods used: operator overloading (e.g. {\tt ADOL-C}, {\tt Adept}) and source code transformation (e.g. {\tt Tapenade}).
\par Source code transformation consists in automatically adding in the source code, before compilation, the instructions to compute the derivatives. It requires a tool able to read an existing source code and to understand it (like a compiler does). Therefore, this method is quite hard to implement. It is often unable to handle advanced features of a language (such as object-oriented programming).
\par Using operator overloading is much easier. Operator overloading is a feature of some programming languages, allowing to redefine basic operators (such as \verb!+!, \verb!*!, \verb!/!). For example, if there is an addition \verb!a+b! in the source code, it will call a customized implementation of \verb!+! (written by the programmer) on the arguments \verb!a! and \verb!b!, instead of performing a traditional addition. In the case of AAD, we redefine the basic operators to also handle the computation of derivatives. It allows to quickly apply AAD to an existing code, with very few modifications to the code. In comparison with source code transformation, operator overloading cannot be as well optimized by the compiler, and may add some overhead during execution. Thus, it is considered as less efficient.
\par For the sake of simplicity, we choose to use operator overloading in {\tt C++}. The specificities of our implementation are that it focuses on Monte Carlo computations, and that it is efficiently parallelized, whether on CPU (using {\tt OpenMP}) or on GPU (using {\tt CUDA}).
 
\subsection{Principle of AAD and basic implementation}
\par We model a computation as a sequence of $n$ intermediate results $x_i$ produced by elementary functions $f_i$: $x_i=f_i(x_1,\dots,x_{i-1})$ for $i=1,\dots,n$. To compute the derivative of the output $x_n$ with respect to the input $x_1$, there are two possible modes of differentiation:
\begin{itemize}
\item The \emph{tangent mode}: we define the tangent of a variable $x$ as $\dot{x}=\frac{\partial x}{\partial x_1}$. We know that $\dot{x}_1=\frac{\partial x_1}{\partial x_1}=1$, so we can compute the tangents from $i=1$ to $i=n$ using the chain rule:
\[\dot{x}_i=\frac{\partial x_i}{\partial x_1}=\frac{\partial}{\partial x_1}f_i(x_1,\dots,x_{i-1})=\sum_{j=1}^{i-1} \frac{\partial f_i}{\partial x_j} \frac{\partial x_j}{\partial x_1}=\sum_{j=1}^{i-1} \frac{\partial f_i}{\partial x_j} \dot{x}_j.\]
Finally, the sensitivity is given by the tangent $\dot{x}_n=\frac{\partial x_n}{\partial x_1}$.
\item The \emph{adjoint mode}: we define the adjoint of a variable $x$ as $\overline{x}=\frac{\partial x_n}{\partial x}$. We know that $\overline{x_n}=\frac{\partial x_n}{\partial x_n}=1$, so we can compute the adjoints from $i=n$ to $i=1$ using the chain rule:
\[\overline{x_i}=\frac{\partial x_n}{\partial x_i}=\sum_{j=i+1}^{n} \frac{\partial x_n}{\partial x_j} \frac{\partial f_j}{\partial x_i}=\sum_{j=i+1}^{n} \overline{x_j} \frac{\partial f_j}{\partial x_i}.\]
Finally, the sensitivity is given by the adjoint $\overline{x_1}=\frac{\partial x_n}{\partial x_1}$.
\end{itemize}
AAD uses adjoint mode: that way, if there are several inputs, all the sentitivities are computed at once (they are given by the corresponding adjoints). That is why it is particularly 
interesting to use AAD in situations where we want to compute sensitivities to many inputs. The downside of AAD is that, since the derivatives are computed in reverse, the values of the $x_i$ must be available before we can start the differentiation.
\par Therefore, to apply AAD in practice, we must first do a forward sweep, where we compute and store every intermediate result $x_i$ from the beginning to the end. We can then do the reverse sweep, where we compute the adjoints $\overline{x_i}$ from the end to the beginning. Here is the resulting algorithm: 

\lstset{mathescape=true,basicstyle=\ttfamily, xleftmargin=\parindent}
\begin{lstlisting}
for $i$ from $1$ to $n$
    $x_i=f_i(x_1,\dots,x_{i-1})$
$\overline{x_i}=0$ for $i<n$, $\overline{x_n}=1$
for $i$ from $n$ to $1$
    for $j$ from $1$ to $i-1$
        $\overline{x_j}\mathrel{{+}{=}} \overline{x_i}\cdot\frac{\partial f_i}{\partial x_j}$
\end{lstlisting}
\par This is very easy to implement using operator overloading: the functions $f_i$ are elementary operators which are overloaded to compute and store the relevant quantities during the forward and reverse sweep. To apply AAD on a Monte Carlo computation, we must differentiate the same computation a large number of times with different inputs (the random variables). Therefore, we apply the algorithm above to vectors: each component of the vectors corresponds to one path of the Monte Carlo. At the end of the computation, we can average the components of any adjoint to get the corresponding sensitivity.

We propose two versions of the algorithm that are very close to each other:
\begin{itemize}
\item A \emph{CPU version}, where the overloaded operators use \verb+for+ loops. To parallelize this version, we use the OpenMP framework, which allows to easily divide the Monte Carlo paths of the \verb+for+ loops among the available CPU cores.
\item A \emph{GPU version}, where the overloaded operators call GPU functions. These functions, also known as kernels, are implemented in CUDA, a proprietary framework for Nvidia graphic cards.
\end{itemize}
\par We need to be careful to avoid useless computations. Some computations are identical across all the Monte Carlo paths: if a computation does not depend on random variables (directly or indirectly), it always has the same inputs and outputs. Therefore, to be efficient, we need to do these computations in a scalar way.
\subsection{Optimisations}
\par
\begin{figure}[!h]
\centering
\includegraphics[page=1,width=0.4592\textwidth]{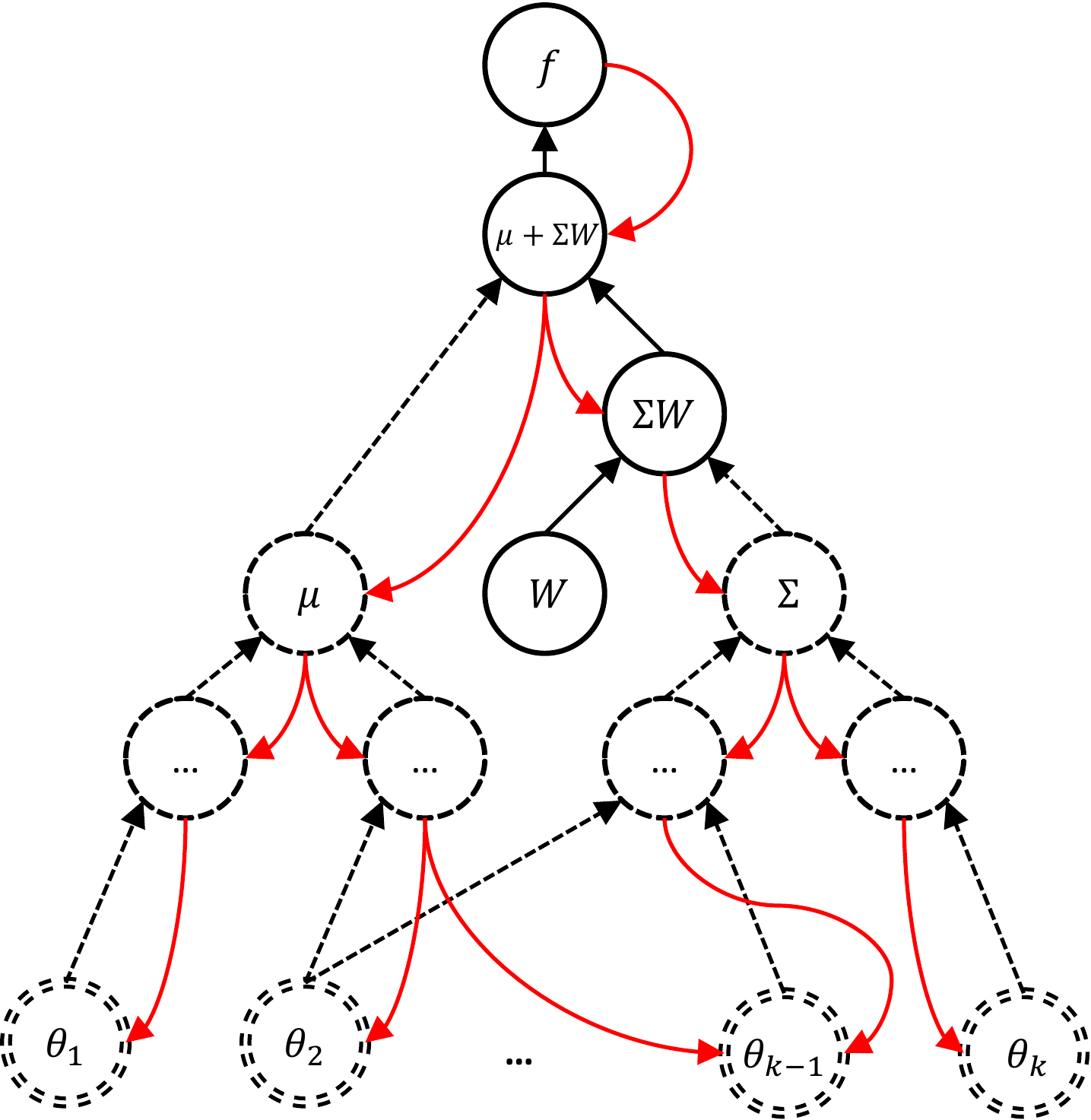}
\hfill
\includegraphics[page=2,width=0.504\textwidth]{graphiques_sg}
\caption{Computational graph for a call on a log-normal underlying, with deterministic drift $\mu(\theta)$ and deterministic volatility $\Sigma(\theta)$. The payoff has the form $f(\theta,W)=\left(\exp\left(\mu(\theta)+\Sigma(\theta)W\right)-K\right)_+$ where $\theta$ represents the inputs and $W$ is a standard gaussian. Each node corresponds to an intermediate result in the forward sweep. Each downward arrow (in red) corresponds to an adjoint update. Solid lines represent vectorial computations and dashed lines represent scalar computations. Means of the adjoints are computed at the doubly circled nodes. On the left (without optimisation), means are computed at the end of the reverse sweep, therefore the reverse sweep is entirely vectorial. On the right (with optimisation), means are computed in the middle of the reverse sweep, therefore the reverse sweep is mostly scalar (which is much cheaper). Here the non-deterministic part is $g(h,W)=\left(\exp\left(h_1+h_2 W\right)-K\right)_+$ and the deterministic part is $h(\theta)=\left(\mu(\theta),\Sigma(\theta)\right)$.}
\label{figure_sg1}
\end{figure}

 Problems of efficiency arise when a big part of the original computation is deterministic. In that case, the forward sweep is cheap, because almost all the computations are scalar. However, the reverse sweep is very expensive, because all the adjoints computations are vectorial (all the intermediate adjoints indirectly depend on the random variables). For example, this can happen for a call on a log-normal underlying, as shown in figure~\ref{figure_sg1} (left). If the deterministic drift and volatility involve a lot of computations, there will be a lot of corresponding adjoint computations (which are expensive because they are vectorial).
\par We can solve this problem by computing intermediate means in the middle of the reverse sweep. If we divide the computation as $f(\theta,W)=g(h(\theta),W)$ where $h=(h_1,\dots, h_q)$ is the deterministic part and $g$ the non-deterministic part, the sensitivity is given by
\[\frac{1}{M} \sum_{i=1}^M \frac{\partial f}{\partial \theta}(\theta, W_i)=\frac{1}{M} \sum_{i=1}^M \sum_{j=1}^q \frac{\partial g}{\partial h_j}(h(\theta), W_i)\frac{\partial h_j}{\partial \theta}(\theta)= \sum_{j=1}^q \left(\frac{1}{M} \sum_{i=1}^M \frac{\partial g}{\partial h_j}(h(\theta), W_i) \vphantom{\frac{1}{M} \sum_{i=1}^M}\right)\frac{\partial h_j}{\partial \theta}(\theta).\]
In practice, it means that during the reverse sweep, we can pause the computation between the deterministic and non-deterministic parts, and replace every (vectorial) adjoint by its (scalar) mean. Thus, when we resume the reverse sweep, the subsequent computations are scalar. The effects of the optimisation for the call are shown on figure~\ref{figure_sg1} (right).
\par Note however that this trick works only because taking the mean of the derivatives is a linear operation. It cannot be applied to a variance computation. But if we want to estimate the variance in addition to the sensitivities, we can divide the Monte Carlo paths into a small number of groups and apply the trick on each group to compute an estimate of the sensitivity; the variance can then be estimated using the different samples of sensitivity obtained on the different groups.
\par To increase the parallelization, we do independent computations in parallel. Two computations are independent if the inputs of one do not depend on the output of the other. Thus, the computations are divided into layers, which are chronologically ordered: each layer can only begin when the previous layer is finished. Inside a layer, everything can be computed in parallel.
\par
\begin{figure}[!h]
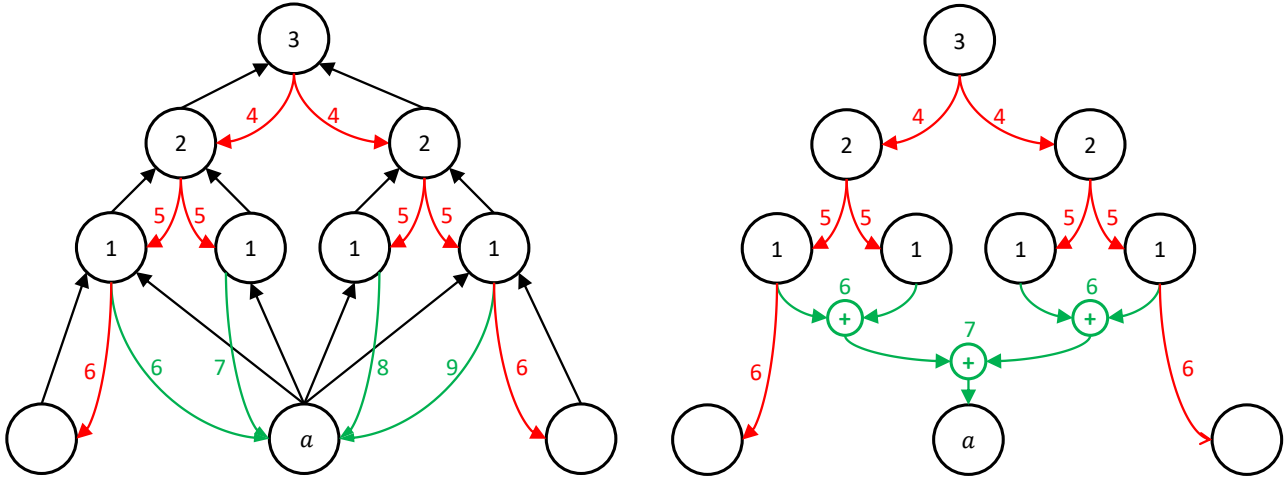

\centering
\includegraphics[page=3,width=0.48\textwidth]{graphiques_sg}
\hfill
\includegraphics[page=4,width=0.48\textwidth]{graphiques_sg}
\caption{Example where an intermediate result (called $a$) is used by many computations. For each node or adjoint update, we have indicated the layer to which the computation belongs. Layers 1 to 3 correspond to the forward sweep. Layers 4 and beyond correspond to the reverse sweep. Without aggregation (on the left), adjoint updates for $a$ (in green) occupy four layers ($6$ to $9$). With aggregation (on the right, where forward arrows have been omitted for the sake of clarity), adjoint updates for $a$ are done with only two layers ($6$ and $7$).}
\label{figure_sg2}
\end{figure}

 When an intermediate result is used as input by many computations, the corresponding adjoint has to be updated many times during the reverse sweep. But since adjoint updates read and modify the adjoints, two adjoint updates on the same variable cannot happen in parallel (they have to be in different layers): if the adjoint updates are done naively (i.e. by adding the updates successively one by one), the number of layers will grow linearly with respect to the number of adjoint updates, which breaks the parallelization. But the resulting adjoint is just a sum of many independent contributions. We can therefore use a divide and conquer method: the big sum can be computed as the addition of two smaller independent sub-sums. Thus, by aggregating the adjoint updates by powers of two, the number of layers grows only logarithmically. This process of aggregation is shown in figure \ref{figure_sg2}.
\par Finally, we can also take advantage of the specificities of the model and the payoff. For example, for a CVA computation, the payoff can be seen as a sum of much simpler payoffs. Therefore, we can replace one big AAD with many small ones, which improves parallelization. Furthermore, we can in the GPU version make the work on CPU (the analysis of dependencies to divide the computational graph into layers) overlap with the work on GPU (the actual computation of intermediate results and adjoints): while the GPU computes the results on one AAD, the CPU works on the next AAD.
\subsection{Results}
\par On a CVA example where we compute 200 sensitivities with 5000 paths, we get the following timings:
\begin{center}
{
\renewcommand{\arraystretch}{1.2}
\begin{tabular}{|l|c|c|}
\cline{2-3}
\multicolumn{1}{c|}{} & CPU & GPU \\ 
\hline 
Finite differences & 271 s & 29 s  \\ 
\hline 
AAD & 9.1 s &  2.4 s \\ 
\hline 
\end{tabular} 
}
\end{center}
We see that when there are many sensitivities to compute, AAD allows interesting gains in comparison with finite differences: a factor of 30 on CPU and of 12 on GPU. It shows that the method can be efficiently parallelized, both on CPU and on GPU, especially if we take advantage of the specificities of the model and payoff. However, AAD cannot be directly applied to non-smooth payoffs. That is why \cite{Gil08} proposed the Vibrato Monte Carlo method, which consists in applying the Likelihood Ratio Method to the conditional expectation at the last step of an Euler scheme.


%
\section{Second Sensitivities in  Quantitative Finance} 
\begin{center}
  {By \emph{Olivier Pironneau}, based on a joint work with  Gilles Pag\`es and Guillaume Sall}
\end{center}

{This short summary presents our work aimed at the computation of higher order sensitivities in quantitative finance; in particular Vibrato and automatic differentiation are compared with other methods. We show that this combined technique is  more stable than automatic differentiation of second order derivatives by finite differences and more general than Malliavin Calculus. We also extend automatic differentiation for second order derivatives of options with non-twice differentiable payoff.}

Sensitivities of large portfolios are time consuming to compute. 
Vibrato \cite{Gil08}, \cite{Gil07} is an efficient method for the differentiation of mean value of functions involving the solution of a stochastic ODE. 
In many cases, double sensitivities, i.e. second derivatives with respect to parameters, are needed (e.g. gamma hedging).  Then one can either apply an automatic differentiation module to Vibrato or try Vibrato of Vibrato, or even do a second order automatic differentiation of the computer program.   But in finance the payoff is never twice differentiable and so generalized derivatives have to be used requiring approximations of Dirac functions of which the precision is also doubtful.

Consider a financial asset $X_t$ modeled by
\[
 {\d X_t = X_t\Big(r(t)\d t + \sigma(X_t,t)\d W_t\Big)},~~~X_0 \hbox{ given},~W_t\hbox{ Brownian}
\]
A European put option
$V_t=e^{-r(T-t)}\E(K-X_T)^+$.
How can we compute the $1^{st}$ and $2^{nd}$ derivatives  of $V_0$ w.r. to $K$,$T$,$X_0$, $r$,$\sigma$. The most straightforward approach is to use the following approximation
\eq{\label{dd2}
\Gamma=\frac{\partial^2 {V_0}}{\partial X_0^2} \approx\frac1{h^2}\Big( V_0|_{X_0+h}- 2 V_0|_{X_0} + V_0|_{X_0-h}\Big)
}
It costs 3 times the computation of the option but it is unstable with respect to $h$:  if $h$ is too small round-off errors polute the result because both the numerator and the denominator are small while the result of the division is not.

One trick is to use a complex $h$. With $\ii:=\sqrt{-1}$,
 \eq{&&
 Re[ \frac{f(a+\ii\delta a)-f(a)}{\ii\delta a}]
 = Im\frac{f(a+\ii\delta a)}{\delta a}=f'(a) - f^{(3)}\frac{\delta a^2}{6} + o(\delta a^3)
}
and it is seen that the expression in the middle is not cursed by a $0$ over $0$ indeterminacy.

\begin{figure}[htbp]
\begin{center}
\includegraphics[width=.32\textwidth]{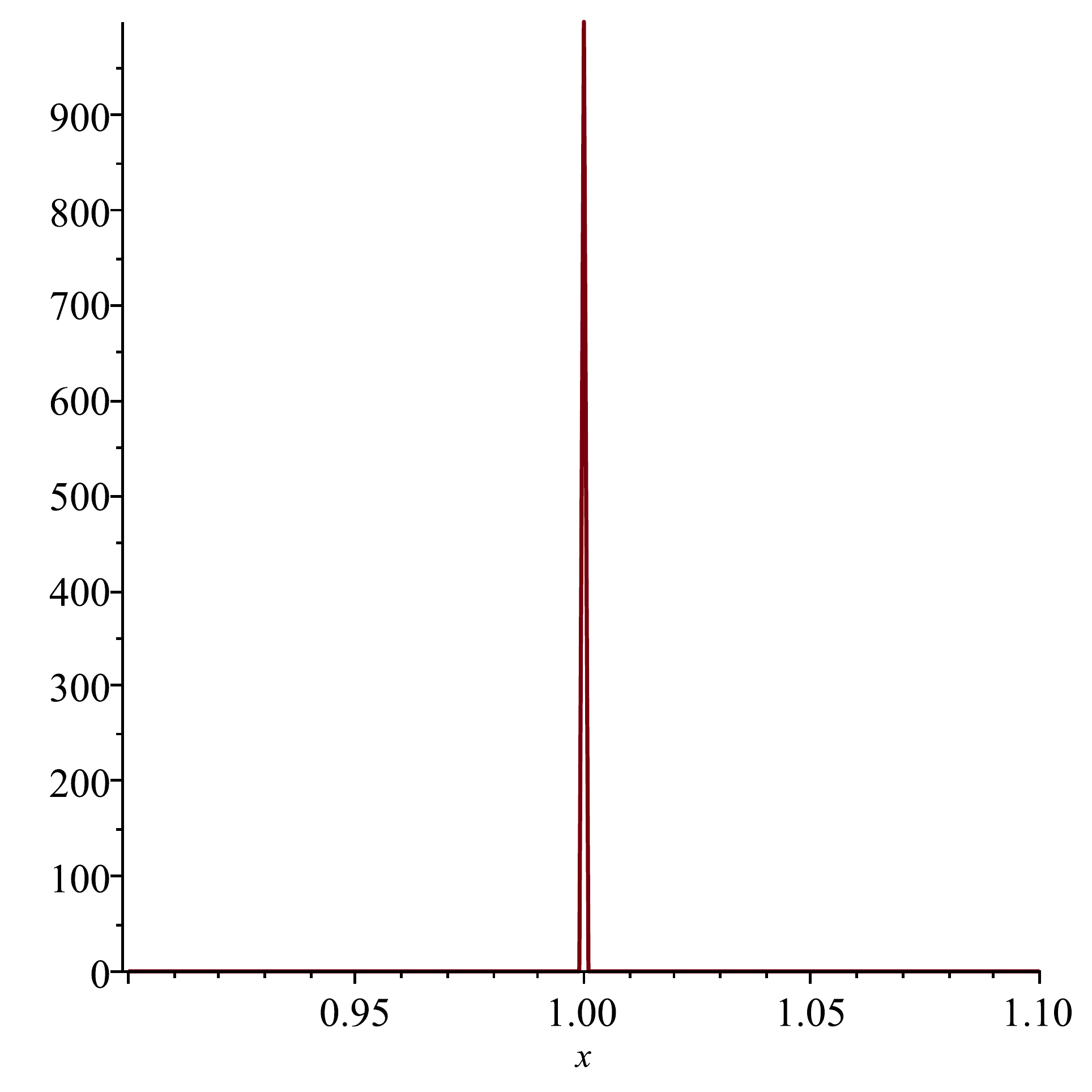}
\includegraphics[width=.32\textwidth]{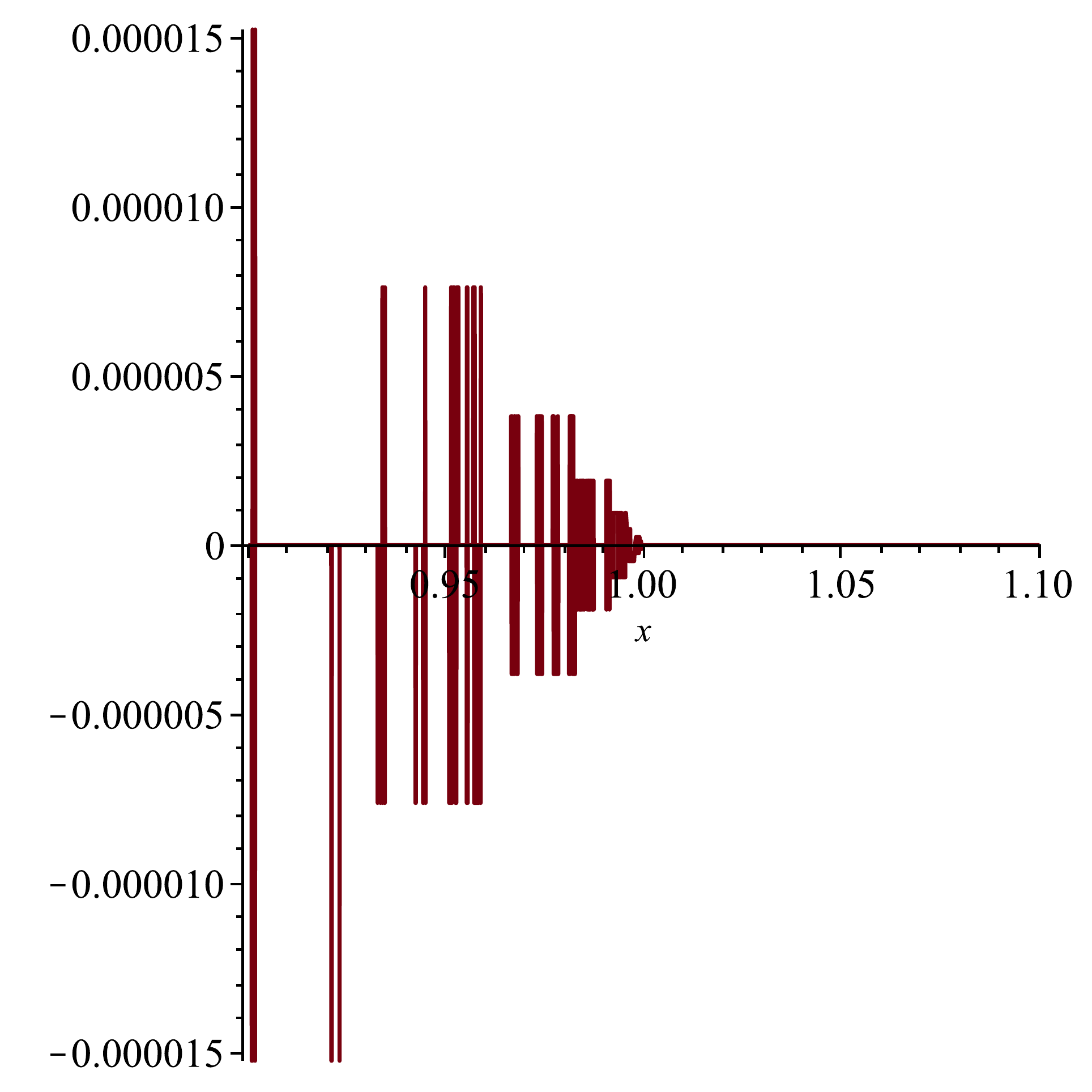}
\includegraphics[width=.32\textwidth]{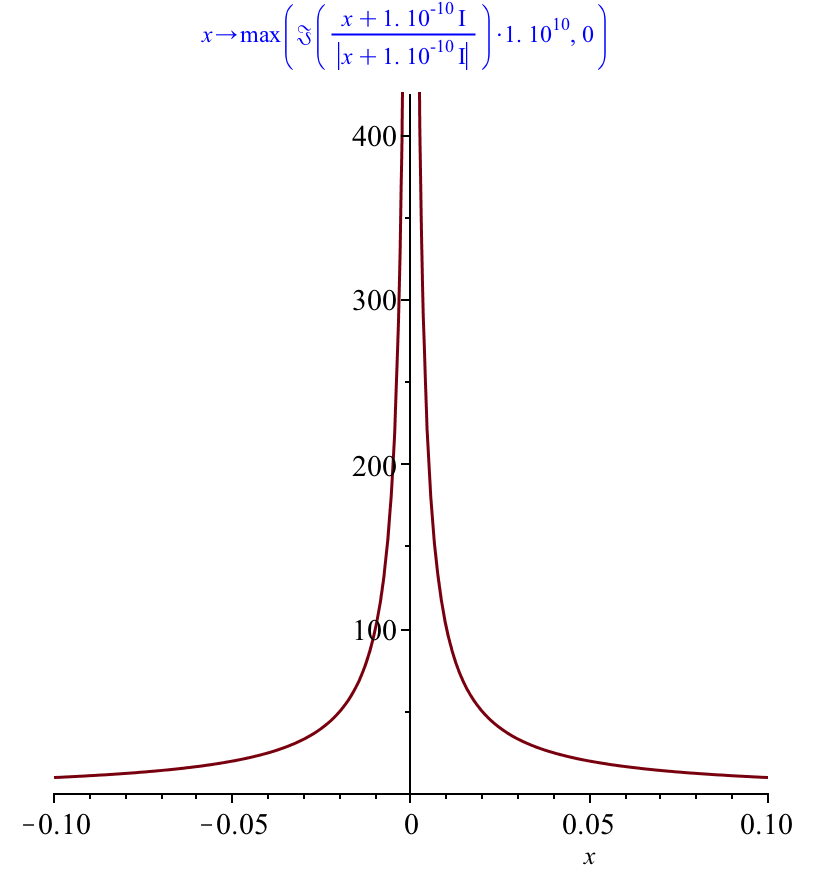}
\caption{using Maple to plot the $2^{nd}$ derivative of  $(1-x)^+$ 
with $\delta x= 10^{-3}$ (left), $\delta x= 3.510^{-6}$ (middle) and $\delta x = \ii 10^{-10}$ (right).}
\label{figtwo}
\end{center}
\end{figure}

Does it  work on non-differentiable functions? For example take $f(x)=(1-x)^+$ leading to $f'(x)=-{\bf 1}_{x<1}$ and $f"(x)=\delta(1-x)$. Using Maple, Figure \ref{figtwo} shows clearly that complex finite difference to compute the second derivative is superior because it gives a correct result even with a very small complex increment.
The conclusion is that unless something is done the results will be unreliable.

 Alternatively Malliavin calculus can be used in some but not all cases.
  For instance, if the payoff $V$ of a vanilla future with spot price $X_t(x)$ where $X_t$ satisfies the Black-Scholes  SDE with constant $\sigma$ and $r$, with $X_0=x$, then $
\Gamma =
\E[{e^{-rT}V(X_T)\over x^2T\sigma }({W_T^2\over\sigma
T}-W_T-{1\over\sigma})], 
$
It works so long as the weight functions are known and $T$ not too small, but when applicable it is clearly faster than anything else.

\subsection{Vibrato}
Let us now turn to Vibrato, as introduced by Giles in \cite{Gil08}.  
Let $\theta$ be a parameter and let us focus on $\p_\theta\E[V(X_T)]$ with
$ dX_t=b\left(\theta, X_t\right)dt +\sigma(\theta, X_t)dW_t,~~X_0=x. 
$ With $Z$ a normal vector valued random variable, an Euler explicit scheme leads to
\begin{equation*}
 	\label{1e} \bar{X}_{{k}}=\bar{X}_{{k-1}}+b(\theta,\bar{X}_{{k-1}})h+\sigma(\theta, \bar{X}_{k-1})\sqrt{h}Z_{k},~~\bar X_0=x,~~k=1,\dots,n.
\end{equation*}
Now write ${\E}\left[V(\bar{X}_{n})\right]={\E}\left[{\E}\left[V(\bar{X}_{n})\mid \bar{X}_{{n-1}}\right]\right]$
and note that, with $\sigma_{n-1}(\theta)=\sigma ( \theta, \bar{X}_{{n-1}}(\theta))$,
\begin{eqnarray}&&
	\label{vb1c} 
	\bar{X}_{n}=\mu_{n-1} (\theta)+\sigma_{n-1}(\theta) Z_n\sqrt h \hbox{ with }
 \mu_{n-1} (\theta)=\bar{X}_{{n-1}}(\theta)+b(\theta, \bar{X}_{{n-1}}(\theta) )h,
\cr&&
\frac{
	\partial}{
	\partial \theta_i}{\E}[V( \bar X_n(\theta) ) ] =\displaystyle {\E_z}\left[ \frac{
	\partial}{
	\partial \theta_i } \left\{{\E}[V(\mu + \sigma Z\sqrt h) ] \right\} _{\tiny 
	\begin{matrix}
		\mu = \mu_{n-1}(\theta) \\
		\sigma = \sigma _{n-1}(\theta) 
	\end{matrix}
	} \right] 
\end{eqnarray}
The last time step sees constant coefficients, so we have an explicit solution. Some algebra 
(see \cite{PPS}) around the identity
	\begin{equation}
		\frac{
		\partial}{
		\partial \theta_i}\Big[ \mathbb{E}[V(X(\theta))]\Big]_{|\theta=\theta^0}=\int_{{\R}^d}V(y) \frac{
		\partial \log{p}}{
		\partial \theta_i}(\theta^0,y)p(\theta^0,y)dy =\mathbb{E}\left[ V(X) \frac{
		\partial \log{p}}{
		\partial \theta_i}\right]_{|\theta^0},
	\end{equation}
where $p$ is the probability density of $X$, leads to the following result.
\begin{proposition}
	\begin{equation*}\label{vib1}
		\begin{aligned}
			\frac{
			\partial }{
			\partial \theta}{\E}[V( \bar X_n(\theta) ) ] &
	={\E}\left[\frac{1}{\sqrt{h}} \frac{
			\partial \mu}{
			\partial \theta} \cdot {\E_z}\left[ V(\mu+\sigma Z\sqrt h)\sigma ^{-T}Z \right] 
			+\frac{1}{2} 
                        \frac{
			\partial (\sigma\sigma^T)}{
			\partial \theta} : {\E_z}\left[ V(\mu+\sigma Z\sqrt h) \sigma ^{-T}( Z Z ^T - I ) \sigma ^{-1} \right] 
                  \right],
		\end{aligned}
	\end{equation*}
where $f\cdot v$ denotes the scalar product between $f$ and $v$ and $M:P$ denotes the trace of the product of 
matrices $M$ and $P$.

\end{proposition}

Note that in the non constant case the tangent process $Y_t=\p_\theta X_t$ is involved.
Note also that $V$ is not differentiated! The computing time in the non constant case is twice the evaluation of $V(X_0)$, similar to finite difference but much more precise.  Note finally that antithetic variance reduction is easy to apply.  More details can be found in \cite{PPS}.
 
\subsection{Higher Derivatives with Automatic Differentiation}
Once the computer program for the vibrato is written, automatic differentiation can be applied. It is exact because every line of the program is differentiated exactly (teaching the compiler for instance by adding in the AD library that the derivative of $\sin x$ is $\cos x$).

Automatic differentiation can be extended to handle the difficulty raised above for non-differentiable pay-off, to some degree, by approximating the Dirac mass at $0$ by 
$
\delta^a(x) = \frac1{a\sqrt{\pi}}e^{-\frac{x^2}{a^2}}.
$
Now, suppose $f$ is discontinuous at $x=z$ and smooth elsewhere; then
$
f(x) = f^+(x)H(x-z)+f^-(x)(1-H(x-z));
$
hence
\[
f'_z(x) = (f^+)'_z(x)H(x-z) + (f^-)'_z(x)(1-H(x-z)) -(f^+(z)-f^-(z))\delta(x-z)
\]
Unless this last term is added, the computation of the second order sensitivities will be wrong.

If in the Automatic  library the ramp function  $x^+$ is defined as $xH(x)$ with its derivative to be $H(x)$, if $H$ is defined with its derivative equal to $\delta^a$ and if in the program which computes the financial asset these new functions are used explicitly, like 
$
(x-K)^+ = \hbox{ramp}(x-K)
$ and not $\max(x-K,0)$, then the second derivative in $K$ computed by the AD library will be $\delta^a(x-K)$. Moreover, it will also compute
\[
\int_0^\infty f(x)(x-K)^+ d x\approx \frac1N\sum _{i=1}^N f(\xi_i)\delta^a(\xi_i-K)
\]
where $\xi_i$ are the $N$ quadrature points of the integral or the Monte$-$Carlo points used by the programmer to approximate the integral.

\subsection{Conceptual algorithm for VAD}\label{BSALGO}
In Algorithm 1, we summarize here what needs to be done to compute a second order sensitivity by VAD with antithetic variance reduction in the general case.
To generate $M$ simulation paths with time step $h=\frac{T}{n}$ of the underlying asset $X$ and its tangent process $\displaystyle Y=\frac{\partial X}{\partial \theta}$ we need $M n$ realisations of a normal random variable $Z$; we need also $M_Z$ realisations of $Z_n^m$, called ${\mathbb Z}_i$, for the last time step for each $m$.  Then given $\bXn_0,\bYn_0$:

\alglanguage{pseudocode}
\begin{algorithm}[h]
\small
\label{GENE} 
  \caption{VAD for second derivatives using antithetic variance reduction}
  \begin{algorithmic}[1]
 \For{ $m=1,..,M$ }
\For {$k=0,\dots,n-2$}  
	\State \(  \bar X^m_{k+1}= \bXn^m_{k}+r h\bXn^m_{k}+\bXn_{k} \sigma\sqrt{h}Z^m_{k+1} \)
	\State \(  \bar Y^m_{k+1}= \bYn^m_{k}+r h\bYn^m_{k}+r'_\theta h\bXn^m_{k}+\left(\bYn^m_{k}
			\sigma+ \sigma'_\theta\bXn^m_{k}\right)\sqrt{h}Z^m_{k+1}
			\)
\EndFor
\label{BSVIB} 
\For {$i=1,..,M_Z$ } 
\label{algoVib2}
\State\( X^m_{T_\pm}=\bXn^m_{n-1}+r h\bXn^m_{n-1}\pm\sigma^m\bXn_{n-1}\sqrt{h}{\mathbb Z}_i,~\bar X_{T_o}=\bXn_{n-1}+r h\bXn_{n-1}\)
\State\( V^m_{T_{\pm,}}=(\bar X^m_{T_{\pm}}-K)^+$, $V^m_{T_o}=(\bar X^m_{T_o}-K)^+.\)
\State\(	R=	\bYn^m_{n-1}(1+r h)+\bXn^m_{n-1}r'_\theta h,~~	S=\bYn^m_{n-1}\sigma+\bXn^m_{n-1}\sigma'_\theta\)		
\State \(  			\ds			\left(\frac{
						\partial V_T}{
						\partial\theta}\right)^m_i=R\frac{(V^m_{T_+}- V^m_{T_-}) {\mathbb Z}_i}{2\bXn^m_{n-1}\sigma \sqrt{h}}
						+S	(V^m_{T_+}-2V^m_{T_o}+V^m_{T_-})\frac{{{\mathbb Z}_i}^2-1}{2\bXn^m_{n-1}\sigma} \)
 \EndFor
\label{BSAUTO} 
\State \( \hbox{Compute $\left(\frac{\partial^2 V_T}{\partial\theta^2}\right)^m_i$ 
by AD on the program that implements $\left(\frac{
						\partial V_T}{
						\partial\theta}\right)^m_i$ }\)
 \label{BSVIBB}
\State \(  \left(\frac{\partial^2 V_T}{\partial\theta^2}\right)^m=\frac1{M_Z}\sum_{i=1}^{M_Z}\left(\frac{\partial^2 V_T}{\partial\theta^2}\right)^m_i\)
 \EndFor
 \State\( \frac{\partial^2 V_T}{\partial\theta^2}=\frac1M\sum_{m=1}^{M}\left(\frac{\partial^2 V_T}{\partial\theta^2}\right)^m.\) 
  \end{algorithmic}
\end{algorithm}
		
		\subsection*{American Option}\label{AMERICAN}
		
The value $V_t$ of an American option requires the best exercise strategy. Let $\varphi$ be the payoff; the dynamic programming approximation of the option is 
\begin{equation}
\bar{V}_{t_n} = e^{-r T} \varphi(\bar X_{T}),~~ C_{t_k}=\mathbb{E}[e^{-r h }\bar V_{t_{k+1}} \mid {\bar X}_{t_k}],~~
				 \bar{V}_{t_k} = \max{\left\{e^{-rt_k}\varphi( \bar X_{t_k} ), C_{t_k}\right\}}.
\end{equation}
Following Longstaff et al. \cite{LS01} the continuation value is approximated by a least square projection on $I$ real valued basis functions, $t_k$, $\{\psi_{i}\}_{i=1}^I$:
		{\small\begin{equation}
			C_{{k}} \simeq \sum_{i=1}^I \alpha _{{k},i}\psi _{i}(\bar X_{{t_k}}),~
		\alpha_{k,\cdot}=\hbox{argmin}_{\alpha}\mathbb{E}\left[\left(e^{-r{h} }\bar V_{t_{k+1}}- \sum_{i=1}^I \alpha _{k,i}\psi _{i}(\bar X_{{k}})\right)^2 \right].
		\end{equation}}
		Once the optimal stopping time $\tau^*$ is known, the differentiation with respect to $\theta$ 
can be done as for a European contract. The dependency of the $\tau ^*$ on $\theta$ is neglected; arguably this dependency is second order but this point needs to be validated.
		
		We consider the following parameters : $\sigma=20\%$ or $\sigma=40\%$, $X_0$ varying from 36 to 44, $T=1$ or $T=2$, $K=40$ and $r=6\%$. The Monte Carlo parameters are: $M=5~10^4$ simulation paths and $T/h=50$ time steps. The basis in the Longstaff-Scharwtz algorithm is $\psi_i(x)=x^{i-1}, i=1,2,3$, i.e. $I=3$.
		
		We compare with the solution of the Black-Scholes partial differential inequation discretized by an implicit Euler scheme in time, finite element in space and semi-smooth Newton for the inequalities  \cite{AP05}. A large number of grid points are used, $10^4$ to make it a reference solution. 
		
		A second order finite Difference approximation is also used to compute the Gamma for comparison. 
		
				
		In Table \ref{amctab}, the results are shown for different sets of parameters taken from Longstaff et al. \cite{LS01}. The method provides a good precision when  variance reduction is used, except when the underlying asset price is small with a small volatility. As for the computation time, the method is faster than Finite Difference.
\begin{table}[!h] 
			\caption{{\small\label{amctab} Results of the price, the Delta and the Gamma of an American option. The reference values are obtained via the Semi-Newton method plus Finite Difference, they are compared to Vibrato plus Automatic Differentiation on the Longstaff-Schwartz algorithm. We compute the standard error for each American Monte Carlo results. The settings of the American Monte Carlo are $50$ time steps and $50,000$ simulation paths.}}
			
			\bigskip \centering\small\setlength\tabcolsep{3pt}
			
			\hspace*{0cm}
{\tiny			\begin{tabular}
				{c c c | c c c c c c c c c } \toprule ~$S$ ~& ~~$\sigma$~ & ~$T$ ~~& 
				\begin{tabular}
					[x]{@{}c@{}}Price\\(AMC)
				\end{tabular}
				& 
				\begin{tabular}
					[x]{@{}c@{}}Standard\\Error
				\end{tabular}
				& ~~ &
				\begin{tabular}
					[x]{@{}c@{}}Delta\\Vibrato (AMC)
				\end{tabular}
				&
				\begin{tabular}
					[x]{@{}c@{}}Standard\\Error
				\end{tabular}
				& ~~ & 
				\begin{tabular}
					[x]{@{}c@{}}Gamma\\Ref. Value
				\end{tabular}
				& 
				\begin{tabular}
					[x]{@{}c@{}}Gamma\\VAD (AMC)
				\end{tabular}
				& 
				\begin{tabular}
					[x]{@{}c@{}}Standard\\Error
				\end{tabular}
				\\
				\midrule
				
				36 & 0.2 & 1  & 4.46289 & 0.013 & ~~&0.68123 & 1.820e$-$3&~~&0.08732 & 0.06745 & 6.947e$-$5\\
				36 & 0.2 & 2  & 4.81523 & 0.016& ~~&0.59934 & 1.813e$-$3&~~&0.07381 & 0.06398& 6.846e$-$5 \\
				36 & 0.4 & 1  &7.07985 &0.016 &~~ &0.51187 & 1.674e$-$3&~~&0.03305 & 0.03546& 4.852e$-$5\\
				36 & 0.4 & 2  &8.45612 &0.024 &~~ &0.44102& 1.488e$-$3 &~~&0.02510 & 0.02591& 5.023e$-$5\\
				
				& & & & & & & &  & &  \\
				
				38 & 0.2 & 1 &  3.23324&0.013 &~~ & 0.53063 & 1.821e$-$3&~~ &0.07349 & 0.07219 & 1.198e$-$4\\
				38 & 0.2 & 2 &  3.72705&0.015&~~ & 0.46732 & 1.669e$-$3&~~ &0.05907 & 0.05789& 1.111e$-$4\\
				38 & 0.4 & 1 &  6.11209& 0.016&~~ &0.45079 & 1.453e$-$3&~~ &0.02989 & 0.03081& 5.465e$-$5\\
				38 & 0.4 & 2 &  7.61031& 0.025&~~ & 0.39503 & 1.922e$-$3&~~ &0.02233 & 0.02342& 4.827e$-$5\\
				
				& & & & & & & & & &  \\
				
				40 & 0.2 & 1 & 2.30565&0.012&~~& 0.40780& 1.880e$-$3& ~~& 0.06014& 0.05954& 1.213e$-$4 \\
				40 & 0.2 & 2 & 2.86072&0.014&~~& 0.39266& 1.747e$-$3& ~~& 0.04717& 0.04567& 5.175e$-$4\\
				40 & 0.4 & 1 & 5.28741&0.015&~~& 0.39485& 1.629e$-$3& ~~& 0.02689& 0.02798& 1.249e$-$5\\
				40 & 0.4 & 2 & 6.85873&0.026&~~& 0.35446& 1.416e$-$3& ~~& 0.01987& 0.02050& 3.989e$-$5\\
				
				& & & & & & & & & &  \\
				
				42 & 0.2 & 1 &  1.60788&0.011&~~& 0.29712& 1.734e$-$3& ~~& 0.04764& 0.04563& 4.797e$-$5\\
				42 & 0.2 & 2 & 2.19079&0.014&~~& 0.28175& 1.601e$-$3& ~~& 0.03749& 0.03601& 5.560e$-$5\\
				42 & 0.4 & 1 & 4.57191&0.015&~~& 0.34385& 1.517e$-$3& ~~& 0.02391& 0.02426&3.194e$-$5\\
				42 & 0.4 & 2 &6.18424&0.023&~~& 0.29943& 1.347e$-$3& ~~& 0.01768& 0.01748& 2.961e$-$5\\
				
				& & & & & & & & & & &  \\
				
				44 & 0.2 & 1 & 1.09648 &0.009&~~& 0.20571& 1.503e$-$3& ~~& 0.03653& 0.03438& 1.486e$-$4\\
				44 & 0.2 & 2 & 1.66903 &0.012&~~& 0.21972& 1.487e$-$3& ~~& 0.02960& 0.02765& 2.363e$-$4\\
				44 & 0.4 & 1 & 3.90838 &0.015&~~& 0.29764& 1.403e$-$3& ~~& 0.02116& 0.02086& 1.274e$-$4\\
				44 & 0.4 & 2 & 5.58252 &0.028&~~& 0.28447& 1.325e$-$3& ~~& 0.01574& 0.01520& 2.162e$-$4\\
				\bottomrule 
			\end{tabular}
			\hspace*{0cm} 
}\end{table}

\noindent{\bf Conclusion.}
Faced with the problem of writing a general software for the computations of sensitivites in connection with Basel III directives, we found that Automatic Differentiation alone is not up to the task, but applied to Vibrato it is dependable and fairly general.


%
\section{CVA with Greeks and AAD}
\label{sec:adil}
\centerline{by \emph{Adil Reghai}}

\subsection{Framework of application}
In \cite{reg15risk}, we proposed a Monte Carlo approach for pricing CVA using the duality relationship between parameter and hedging sensitivities as well as Ito integral calculus.

CVA is the market-value of counterparty credit risk. It is computed as the difference between the risk-free value of the portfolio and its true value that takes into account counterparties default. When both parties to a bilateral contract may default, the price adjustment is called BCVA (Bilateral Credit Valuation Adjustment).

Under the Monte Carlo approach, CVA is rewritten as a weighted sum of expected discounted exposures from time of calculation $t$ to the maturity of the portfolio $T$. A number of paths for portfolio underlying market variables are simulated and exposures distributions are obtained at each time between t and T. Expected exposure is then given by averaging exposures along simulated paths. Although, Monte Carlo simulations are quite easy to implement, for some complex portfolios this tool could easily lead to heavy computational requirements, thus quickly proving unworkable. Recently,\cite{bur10pderisk} introduced a PDE representation for the value of financial derivatives under counterparty credit risk. Using the Feynman-Kac theorem, the total value of the derivative is decomposed into a default free value plus a second term corresponding to the CVA adjustment. However, for dimensions higher than 3, the PDE can no longer be solved with a finite-difference scheme. Building on this work, \cite{lab12branch} introduces a new method based on branching diffusions using a Galton-Watson random tree. Although this method drastically reduces computational costs, it is still complex to implement. 

Summarizing, we see that both existing approaches for CVA computation suffer from serious drawbacks. Hence we introduced a new approach for CVA valuation in a Monte Carlo setting. Our approach is innovative in that it combines both the martingale representation theorem and Adjoint Algorithmic Differentiation (AAD) techniques to retrieve future prices in a highly efficient manner.

To give a rough understanding of the methodologie, that is more ``trajectory by trajectory''-driven than the two previously existing ones (Monte-Carlo or PDE), it is needed to set few notations:
\begin{itemize}
\item The price of an equity follows here a geometric Brownian motion with a drift made of the risk free rate $r_t$ and a financing cost $\phi_t$:
  $${dS_t \over S_t} = (r_t + \phi_t)\, dt + \sigma_t \, dW_t.$$
  Assuming the risk free rate is zero for the sake of simplicity, any payoff $V(t,S_t)$ of a derivative with a maturity $T$ written on such an equity hence follows the Back-Scholes backward PDE:
  \begin{equation}
    \label{eq:VBS}\nonumber
      {\partial V \over \partial t} + S_t\, \phi(t, S_t)\, {\partial V\over \partial S} + {1\over 2}\, \sigma^2(t,S_t)\, S^2_t \; \frac{\partial^2 V}{\partial^2 S} = 0.
  \end{equation}
\item In the scope of this focus on a use of AAD in finance, we will consider a case in which, seen from the issuer, only the default of the counterparty can occur, and take as given the expression of the Credit Valuation Adjustment (CVA)  in such a context: a call option with zero strike on the net present value of a ``portfolio'' of one asset at the random time of default of the counterparty. Following \cite{brig08pde}, define CVA as
  $$\CVA=\LGD \cdot \Esp\left(\one_{\tau\leq \tau}\, D(0, \tau)\, X(\tau)\right),$$
  where: LGD is the ``Loss Given Default'' (i.e. the fraction of loss incurred by the investor upon default of his counterparty), $\tau$ is the default time, $D(u,v)$ is the discount factor between $u$ and $v$ (under our zero risk free rate assumption: $D(u,v)=1$), and $X(t)$ is the ``exposure\footnote{We furthermore assume the exposure $X$ and the default of the counterparty are independant.} at time $t$''. 
  $X(t)$ can hence be written as the positive part of a solution $V$ of (\ref{eq:VBS}).
\item Last but not least, we assume here the hazard rate $\lambda$ is constant and known, meaning the probability to default reads $\mathbb{Q}(\tau\leq t)=1-e^{ -\lambda t}$. 
\item All this allows to write the CVA seen from time $t\leq\tau$ (\cite[Section 3]{reg15risk} for details) as 
  \begin{equation}
    \label{eq:CVAcont}
      \CVA_t = \Esp_t \int_{u=t}^\tau \LGD\cdot V(u)^+ \lambda e^{-\lambda (u-t)}\, du.
  \end{equation}
  A way to enable the use of AAD here is to consider a pathwise version of $\Esp_t$ (the expectation seen from $t$) using $\Np$ paths and to write this approximation of the $\CVA_t$ (with the notation $v_i$ for the value along a given trajectory, i.e. the cost of this trajectory):
  \begin{equation}
    \label{eq:CVAt}
    \CVA_t = {1\over \Np} \sum_{i=1}^\Np \underbrace{\int_{u=t}^\tau \LGD\cdot v_i(u) \cdot \one_{v_i(u)\geq 0}\;\lambda\, e^{-\lambda (u-t)}\, du}_{\CVA_t^i}.
  \end{equation}
\end{itemize}

In \cite[Section 4]{reg15risk}, it is first established the following relationship
$$\left.{\partial V\over \partial \phi(u,S^i_u)}\right|_{(t,S_t)}=\Esp_t \int_{z=t}^T \one_{(z=u,S_z=S_u^i)} S_u^i\, {\partial V\over \partial S}(z,S_z)\, dz.$$
It means the partial derivative of the value function with respect to the financial costs at any time and price $(u,S^i_u)$, can be expressed as the conditional expectation of the sum the price $S_u^i$ times the delta of the value function $V$ sampled when $z=u$ and $S_z=S_u^i$.
Keep in mind any point of the surface $(u,S^i_u)\mapsto \phi(u,S^i_u)$ can be chosen to compute the partial derivative of $V(t)$, and hence the ``sampling'' at $z=u$ and $S_z=S_u^i$ means any trajectory crossing $(u,S^i_u)$ at $z$ in $[t,T]$ is considered. The mass of these events being $f_S(u,S_u^i):=\int_{z=t}^T \one_{(z=u,S_z=S_u^i)}\, dz$, we can then write
\begin{equation}
  \label{eq:Vstar}
   \underbrace{{\partial V\over \partial S}(u,S_u^i)}_{\mbox{hedging sensitivity}} = \frac{1}{S^i_u\, f_S(u,S_u^i)}\cdot \underbrace{\partial V\over \partial \phi(u,S^i_u)}_{\mbox{input sensitivity}}.
\end{equation}

Then, AAD techniques come into play in order to compute hedging sensitivities by applying the chain rule of instructions in a reversed order with regards to its original formulation. As shown in \cite{aad11capri}, the execution time of the AAD code is bounded by approximately four times the cost of execution of the original function.\medskip

Our approach to compute CVA could be summarized by the following four steps procedure:
\begin{enumerate} 
\item[1.] {\bf Solve the underlying value function $V$.}\\
  Generate a first Monte Carlo sapling consisting of $N^{(1)}_{\rm paths}$ for the stock process. In this first sampling phase we determine the initial price $V(t)$ (price at time \redok{$t$} of CVA valuation) as well as all input sensitivities ${\partial V\over \partial \phi}(t,S)$ using the chain rule of AAD.\medskip
\item[2.] {\bf Use the knowledge of $V$ to sample the input sensitivities.}\\
  Generate a second Monte Carlo sample consisting of $N^{(2)}_{\rm paths}$ for the stock process. At each node (path $i$, time $u$) of the stock price: 
  \begin{itemize}
  \item Get corresponding input sensitivities ${\partial V\over \partial \phi} (u, S^i_u)$ by linear interpolation using results in step 1.
  \item Calculate corresponding hedging sensitivities using (\ref{eq:Vstar}).
  \item Calculate $v_i(u)$ for each trajectory $i$ at each time $u$ (thanks to the martingale representation theorem allowing to express $v_i$ as a perturbation of $V$).
  \end{itemize}\medskip
\item[3.] {\bf Pathwise CVA estimation.}\\
  Now it is possible to compute $\CVA_t^i$ along each path using its expression in formula (\ref{eq:CVAt}).\medskip
\item[4.] {\bf Averaging} these $\CVA_t^i$ over the $N^{(2)}_{\rm paths}$ paths leads to an estimate of the CVA defined by (\ref{eq:CVAcont}).
\end{enumerate}\medskip

Note the use of AAD unlocks step 1. Without such a systematic an automated method, it would be needed to go back to a finite difference approach, having no gain in using formula (\ref{eq:Vstar}).

Moreover, it is possible to extend this AAD based approach to several equity payoffs and even to some path dependent ones \cite{reg15risk}.

\subsection{Practical relevance of our research paper}

The method we presented has potentially a considerable impact on the way banks calculate and manage their CVA. Indeed, not only AAD methods enhance the computational burden of CVA, but they also allow one to obtain sensitivities with regards to all input parameters (in a row). This turns out very crucial for all banks that need to risk manage their derivatives portfolios and to fullfil regulatory demand (see the Introduction of paper).

\begin{table}[h]
  \centering
  \begin{tabular}{|c|c|c|} \hline
    Method & Number of paths & Computing time \\\hline\hline
    CVA with MC-MC & MC for prices: 10,000  & 5 hours \\
                              & MC for CVA: 10,000    & \\\hline
     CVA with AAD    & Step 1: 1,000,000 & 29 seconds\\
                             & Step 2: 10,000  & \\\hline
  \end{tabular}
  \caption{Comparison of a standard approach (MC-MC: twice Monte-Carlo), with the proposed AAD one: two steps two, but automatic differentiation.}\label{tab:ar1}
\end{table}

As far as CVA is concerned, we tested our approach for a number of equity payoffs and compared it to more standard approaches (basically MC of MC to calculate exposures at future times). As a matter of fact, we compared the computing time  required to value CVA for a double barrier option (2 barriers: up and down) where no closed form solution for neither price nor CVA exists. For this option, the price is computed using Monte Carlo and CVA using a Monte Carlo of Monte Carlo (simulation of simulation). The main results are summarized in Table \ref{tab:ar1}.

Results have been obtained with R code with a Processor Intel({\sc r}) Core({\sc tm}) i7-2600 CPU @ 3.40 GHz  and 16 GB of installed RAM. So, we see that the speed up in computing time is over 620 times. Moreover, we have directly the surface of hedging sensitivities using the AAD method (with no other simulations required), whereas simulations would be needed to value hedging sensitivities in the standard MC-MC framework.


\noindent{\bf Disclaimer.}
Views expressed by authors are the authors' own and do not necessarily reflect the view of the companies they work at.


\end{document}